\documentclass[epj]{svjour}
\usepackage{epsfig,bm,amssymb}
\twocolumn

\newcommand{\vep}{{\bm p}}
\newcommand{\vek}{{\bm k}}
\newcommand{\veq}{{\bm q}}
\newcommand{\be}{\begin{equation}}
\newcommand{\ee}{\end{equation}}

\begin{document}

\title{Interplay of quark and meson degrees of freedom in a near-threshold resonance}
\author{V. Baru\inst{1,2}, C. Hanhart\inst{2}, Yu. S. Kalashnikova\inst{1}, A. E. Kudryavtsev\inst{1}, 
A. V. Nefediev\inst{1}}
\institute{Institute for Theoretical and Experimental Physics,
117218, B. Cheremushkinskaya 25, Moscow, Russia 
\and
Forschungszentrum J\"ulich, Institut f\"ur Kernphysik (Theorie) and J\"ulich Center for Hadron
Physics, D-52425 J\"ulich, Germany}

\date{}
\headnote{\hfill\normalsize\rm FZJ--IKP(TH)--2009--39}

\abstract{
We investigate the interplay of quark and meson degrees of freedom in a physical state representing a 
near-threshold resonance for the case of a single continuum channel. We demonstrate that such a 
near-threshold resonance may possess quite peculiar properties if both quark and meson dynamics generate 
weakly coupled near-threshold poles in the S-matrix. In particular, the scattering t-matrix may possess 
zeros in this case. We also discuss possible implications for production reactions as well as studies within lattice QCD. 
}                  
\authorrunning{V. Baru {\em et al.}}
\titlerunning{Interplay of quark and meson degrees of freedom in a 
near-threshold resonance}
\maketitle

\section{Introduction}

With recent developments of $B$-factories, a wealth of new  
charmonia states was reported, with properties of some of those 
being incompatible with simple quark model predictions. 
Exotic explanations, such as hybrids or tetraquarks, are 
suggested to describe the new states. On the other hand, these new 
enigmatic states are above open charm threshold, so that the 
spectrum of the ``quenched" quark model is to be significantly distorted 
by charmed-meson pairs (for reviews, see e.g.
Refs.~\cite{charmreview1,charmreview2,charmreview3,charmreview4,charmreview5}).

Moreover, some of the new states reside in the vicinity of $S$-wave
thresholds. A most prominent example here is the
famous $X(3872)$ state, first observed in $B$-meson decays 
\cite{Xobservation}. The $X(3872)$ is extremely close to the 
$D\bar{D}^*$ threshold, with $D \bar{D}^*$ being in the $S$-wave, if 
the quantum numbers of the $X$ are $1^{++}$, as suggested by the data.
There are also the $Y(4260)$ and $Y(4325)$ vector states
\cite{Yobservation1,Yobservation2}, with relevant $S$-wave thresholds being
the $D \bar{D}_1$ at $4.285$~GeV and $D^*\bar{D}_0$ at about 
$4.360$~GeV ($D_1$ and $D_0$ are correspondingly $1^+$ and $0^+$ 
$D$-mesons). 
The purely exotic charged state $Z^+(4430)$ \cite{Zobservation}
is close to the $D^* \bar{D}_1$ threshold
and, for the charged $Z^+_1(4050)$ and $Z^+_2(4250)$ 
\cite{Z12observation}, the relevant thresholds are 
$D^* \bar{D}^*$ and $D_1 \bar{D}$, respectively\footnote{These charged states 
are to be considered with caution, as seen only in one experiment 
(Belle). Besides, as quantum numbers of the $Z$-particles are not 
known, the thresholds are not necessarily the $S$-wave ones.}. 

Threshold proximity implies that, independently of the binding 
mechanism, there should be a significant 
component of a hadronic molecule in the wave function of the state. So 
the question arises
of how to distinguish between a genuine ``elementary" particle 
($q \bar{q}$, hybrid, or compact tetraquark) and a composite 
state (hadronic molecule), and how to estimate the admixture of the
latter. 
It was suggested in Ref.~\cite{Weinberg1,Weinberg2,Weinberg3} that it is possible, 
in the case of a near-threshold bound state, to answer 
this question in a model-indepen\-dent way: the state is mostly 
elementary if the effective radius is large and negative. The 
approach was generalised in Ref.~\cite{evi1,evi2} to the case of presence of inelastic
channels, as well as to the case of an above-threshold resonance. Related to 
this is the pole counting approach \cite{Morgan}, in which the structure of the
near-threshold
singularities of the scattering amplitude is studied. It appears 
that the state is mostly elementary if there are two nearby poles 
in the scattering amplitude, while a composite particle corresponds 
to a single near-threshold pole.

The approaches of Ref.~\cite{Weinberg1,Weinberg2,Weinberg3,evi1,evi2,Morgan} 
are based on the ef\-fec\-ti\-ve-range expansion of the scattering 
amplitude and, as such, are expected to be valid for the momenta 
involved much smaller than the inverse range of the force. However, as
noticed in Ref.~\cite{Weinberg1,Weinberg2,Weinberg3}, the effective-range formulae are 
not the most general ones even in the small-momenta limit. The 
scattering amplitude, as a function of energy, can have a zero
and, if this zero is situated in the near-threshold region, the
effective-range expansion would fail. In the present paper we 
develop a formalism which allows one to pinpoint the source of this 
failure, to study the hadronic observables in the presence of this 
zero, and to identify the physical situation in which it occurs.  

\section{General formalism}

We consider a physical state which is a mixture of a bare state
($q \bar q$ or compact tetraquark) 
and a dynamical
(mo\-le\-cu\-le) component and represent its wave function as:
\be
|\Psi\rangle=\left(c|\psi_0\rangle \atop \chi|M_1M_2 \rangle\right),
\label{state}
\ee
where $|\psi_0\rangle$ is the bare elementary state with the 
probability 
amplitude $c$, while $\chi(\vep)$ describes the relative motion in 
the system of two mesons ($M_1M_2$), with the masses $m_1$ and $m_2$,
respectively, and with the relative momentum $\vep$. The wave function $|\Psi \rangle$
obeys a Schr{\"o}dinger-like equation:
\be
{\cal H}|\Psi \rangle=E|\Psi \rangle,
\label{Sheq}
\ee
with the Hamiltonian
\be
{\cal H}=
\left(
\begin{array}{cc}
H_0&V_{qh}\\
V_{hq}&H_h
\end{array}
\right),
\label{H}
\ee
where 
\be
H_0|\psi_0\rangle=E_0|\psi_0\rangle, 
\ee
$m_1+m_2+E_0$ is the bare state mass, and
\be
H_h(\vep,\vep')=\frac{p^2}{2\mu}\delta(\vep-\vep')+V(\vep,\vep'),
\ee
where $\mu$ is the reduced mass. 

The term $V_{qh}$ is responsible for the dressing of the bare 
state, which is given by the transition form factor $f(\vep)$:
\be
\langle \psi_0|V_{qh}|M_1M_2 \rangle=f(\vep).
\ee

The Schr{\"o}dinger-like equation (\ref{Sheq}) is equivalent to the system of
equations for the $c$ and $\chi$, which reads:
\be
\left\{
\begin{array}{l}
\displaystyle c(E)E_0+\int f(\vep)\chi(\vep)d^3p=c(E)E,\\[-2mm]
{}\\[-2mm]
\displaystyle\frac{p^2}{2\mu}\chi(\vep)+c(E)f(\vep)+\int
V(\vep,\vek)\chi(\vek)d^3k=E\chi(\vep).
\end{array}
\right.
\label{system}
\ee

On substituting $c(E)$ from the first equation to the second one, we arrive
at the Schr{\"o}dinger equation in the mesonic channel, with an effective 
potential
\be
V_{\rm eff}(\vep,\vep',E)=V(\vep,\vep')+\frac{f(\vep)f(\vep')}{E-E_0}.
\ee
The off-shell mesonic $t$-matrix $t(\vep,\vep',E)$ is a solution of
the Lippmann--Schwinger equation\footnote{Normalisation of the
$t$-matrix is such that the $M_1M_2$ scattering amplitude is given by
$f(\vek,\vek,E)=-4\pi^2\mu t(\vek,\vek,E)$, with
$E=k^2/(2\mu)$.},
\be
t(\vep,\vep',E)=V_{\rm eff}(\vep,\vep',E)
-\int d^3q 
\frac{V_{\rm eff}(\vep,\veq,E)t(\veq,\vep',E)}{q^2/(2\mu)-E-i0}.
\ee
The solution of this equation can be written as
\be
t(\vep,\vep',E)=t_V(\vep,\vep',E)+
\frac{\phi(\vep,E)\bar{\phi}(\vep',E)}{E-E_0+{\cal G}(E)},
\label{tsol}
\ee
where $t_V(\vep,\vep',E)$ is the $t$-matrix for the potential problem,
\be
t_V(\vep,\vep',E)=V(\vep,\vep')-\int d^3q
\frac{V(\vep,\veq)t_V(\veq,\vep',E)}{q^2/(2\mu)-E-i0},
\ee
while the dressed vertex functions are 
\begin{eqnarray}
\phi(\vep,E)&=&f(\vep)-\int d^3q
\frac{t_V(\vep,\veq,E)f(\veq)}{q^2/(2\mu)-E-i0},
\label{dvf1}\\
\bar{\phi}(\vep,E)&=&f(\vep)-\int d^3q
\frac{t_V(\veq,\vep,E)f(\veq)}{q^2/(2\mu)-E-i0},
\label{dvf2}
\end{eqnarray}
and
\be
{\cal G}(E)=\int d^3q\frac{f(\veq)\phi(\veq,E)}{q^2/(2\mu)-E-i0}.
\ee

The system of equations (\ref{system}) can possess bound states (generally, more than one). 
For a bound state $i$ with the binding energy 
$\epsilon_B^{(i)}$, the solution of the system (\ref{system}) may be written as
\be
c_B^{(i)}=\cos\theta_i,\quad \chi_B^{(i)}(\vep)=\psi_i(\vep)\sin\theta_i,
\ee
where $\psi_i(\vep)$ is normalised to unity. Then
the wave function of this bound state,
\be
|\Psi\rangle_B^{(i)}=\left(\cos\theta_i|\psi_0\rangle\atop
\sin\theta_i\;\psi_i(\vep)|M_1M_2\rangle\right),
\label{boundstate}
\ee
is also normalised, ${}_B^{(i)}\langle\Psi|\Psi\rangle_B^{(i)}=1$.

The quantity
\be
Z_i=|\langle\psi_0|\Psi\rangle_B^{(i)}|^2=\cos^2\theta_i,
\label{Zf}
\ee
introduced in Ref.~\cite{Weinberg1,Weinberg2,Weinberg3}, gives the probability to find a bare state
in the wave function of the bound state $i$. 

The solution of the system (\ref{system}) for the continuum is
\be
c_k(E)=\frac{1}{E-E_0}\int  d^3p\chi_k(\vep)f(\vep)=
\frac{\bar{\phi}(\vek,E)}{E-E_0+{\cal G}(E)},
\label{c}
\ee
\be
\chi_k(\vep)=\delta(\vep -\vek)-\frac{t(\vep,\vek,E)}
{p^2/(2\mu)-E-i0},\quad E=\frac{k^2}{2\mu},
\label{chi0}
\ee
so that the continuum counterpart of the quantity (\ref{Zf}), the spectral 
density $w(E)$, gives the probability to find the bare state in 
the continuum wave function \cite{bhm}. It can be found as:
\be
w(E)=4\pi\mu k|c_k(E)|^2\Theta(E),
\label{wE}
\ee
with $c_k(E)$ given by Eq.~(\ref{c}).
As shown in Ref.~\cite{bhm}, the normalisation condition for the 
distribution $w(E)$ reads:
\be
\int_0^{\infty} w(E)dE=1-\sum_i Z_i,
\label{wnorm}
\ee
where the sum goes over all bound states present in the system, while the
corresponding $Z$-factors are given by Eq.~(\ref{Zf}).

We are interested in the energy range close to the threshold.
To perform a low-energy reduction, let us assume that 
the scattering length approximation is valid for the potential problem,
so that the potential $t$-matrix $t_V$ takes the form:
\be
t_V(\veq,\vep,E)=-\frac{1}{4\pi^2\mu(-a_V^{-1}-ik)}+\ldots,
\ee
where $a_V$ is the scattering length for the potential $V(\vep,\vep')$
and ellipsis stands for the range of 
forces corrections. We use the sign convention,
where a negative scattering length corresponds to an attractive 
potential, however, too weak to produce a bound state while a positive
scattering length reverses to either a repulsive potential (which cannot produce a singularity of
the $S$-matrix, thus we automatically
have $1/a_V\sim \beta$, $\beta$ is the range of forces) or a bound 
state, which is the case of interest here.
Note, the above expression is
useful only for $a_V\beta\gg 1$ and then applicable only for momenta
of order $1/a_V$. In this case $t_V$ has a near-threshold pole. However,
it should be stressed that this singularity is observable only in the weak
coupling limit to the quark states located nearby, for otherwise all singularities
undergo a significant mixing, the origin of which is the appearance
of $t_V$ in the dressed vertex functions of Eqs.~(\ref{dvf1}) and (\ref{dvf2}).
For a detailed discussion of this effect see Ref.~\cite{mitmichael1,mitmichael2}.

It is convenient to define the loop functions:
\begin{eqnarray}
g(E)=\int d^3q\frac{f^2(\veq)}{q^2/(2\mu)-E-i0}=f_0^2(R+4\pi^2\mu ik),
\nonumber\\[-2mm]
\\[-2mm]
g'(E)=\int d^3q\frac{f(\veq)}{q^2/(2\mu)-E-i0}=f_0(R'+4\pi^2\mu ik),
\nonumber
\end{eqnarray}
where the constants $R$ and $R'$ are expected to be of order $\mu\beta$
and  corrections $O(k^2/\beta^2)$ where dropped; in addition we
introduced $f_0\equiv f(0)$.
Defining 
\be
R_V=\frac{4\pi^2\mu}{a_V},
\ee
one arrives at the following form for the on-shell $t$-matrix (\ref{tsol}):
\be
t(E)=\frac{E-E_C}{[E-E_0]R_V+f_0^2[RR_V-R'^2]
+4\pi^2\mu ik[E-E_C]},
\label{tmatrix}
\ee
$$
E_C=E_0-f_0^2(R+R_V-2R').
$$ 
Expression (\ref{tmatrix}) for the scattering amplitude has a 
zero at $E=E_C$, and the effective-range expansion for the amplitude
(\ref{tmatrix}) is not valid, if $E_C$ is in the near-threshold 
region.

\section{Flatt{\'e} formulae}

In this chapter we develop a generalisation of the well-known
Flatt{\'e} parameterisation \cite{flatte} for the near-threshold scattering 
amplitude, and  
express the $t$-matrix of Eq.~(\ref{tmatrix}) in terms of Flatt{\'e}-type parameters. 

Let us define the Flatt{\'e} energy $E_f$ such that the real part of the denominator 
in Eq.~(\ref{tmatrix}) has a zero when $E=E_f$,
\be
[E_f-E_0]R_V+f_0^2[RR_V-R'^2]=0,
\ee
and get rid of $E_0$ this way. Then one can
express $E_C$ in terms of $E_f$:
\be
E_C=E_f-\frac{f_0^2}{R_V}(R'-R_V)^2,
\label{ECdef}
\ee
and, using the identity
\be
\frac{E-E_f}{E-E_C}=\frac{E-E_f}{E_f-E_C}-
\frac{(E-E_f)^2}{(E-E_C)(E_f-E_C)},
\ee
rewrite the $t$-matrix in the Flatt{\'e}-type form as:
\be
t(E)=\frac{g_f}{8\pi^2\mu {\cal D}_F(E)}.
\label{tf}
\ee
Here
\be
\frac{g_f}{2}=4\pi^2\mu\frac{E_f-E_C}{R_V}=\frac{4\pi^2\mu f_0^2}{R_V^2}
(R-R_V)^2
\label{gf}
\ee
and
\be
{\cal D}_F(E)=E-E_f-\frac{(E-E_f)^2}{E-E_C}+\frac{i}{2}g_f k.
\label{Fden}
\ee
If $|E_C| \gg |E_f|$, we are back to the standard Flatt{\'e} 
approximation 
for the near-threshold scattering amplitude \cite{flatte}, 
\be
t_F(E)=\frac{1}{8\pi^2\mu}\frac{g_f}{E-E_f+\frac{i}{2}g_f k},
\ee
which depends on two parameters, $g_f$ and $E_f$, in contrast to 
the expression (\ref{tf}), which depends on three parameters: 
$g_f$, $E_f$, and $E_C$.
One can see that, for $|E_C|\sim|E_f|$, the standard Flatt{\'e} formula is 
severely distorted by the presence of the $t$-matrix zero, with 
a drastic effect on the behaviour of the elastic scattering cross 
section.

The Flatt{\'e}-type formulae given above define the $t$-matrix in terms of 
the parameters
$E_f$, $g_f$, and $E_C$. Following S. Weinberg \cite{Weinberg1,Weinberg2,Weinberg3}, it is
instructive to consider the case of a single bound state present, and to 
express the $t$-matrix in
terms of the binding energy $\epsilon_B$, the $Z$-factor (which is 
the probability to find a bare state in the physical bound state 
wave function), and the scattering length $a_V$.

We start from the expression for the inverse $t$-matrix (\ref{tmatrix}),
\be
t^{-1}(E)=\frac{E-E_f}{E-E_C}R_V+4\pi^2\mu ik.
\label{tmin1}
\ee

In the vicinity of the bound-state pole one has
\be
t^{-1}(E)\simeq\frac{\epsilon_B +E}{g_{\rm eff}^2},
\ee
with (see Ref.~\cite{Weinberg1,Weinberg2,Weinberg3})
\be
g_{\rm eff}^2=\frac{\sqrt{2\mu\varepsilon_B}}{4\pi^2\mu^2}(1-Z).
\ee 
This defines
\be
\frac{Z}{1-Z}=\frac{2\epsilon_B}{\epsilon_B +E_C}
\left(1-\frac{\gamma_V}{\sqrt{2\mu\varepsilon_B}}\right),
\label{Z}
\ee
with $\gamma_V=1/a_V$ being the inverse scattering length for the 
potential problem in the absence of the quark state.
Thus the following formulae can be obtained \cite{Weinberg1,Weinberg2,Weinberg3}:
\be
k\cot \delta = 
-\sqrt{2\mu\varepsilon_B}+\frac{\sqrt{2\mu\varepsilon_B}(E+\epsilon_B)
(\epsilon_B + E_C)}
{2\epsilon_B(E-E_C)}\frac{Z}{1-Z},
\label{kcot}
\ee
and
\be
E_C=-\epsilon_B
\left(1-\frac{2(1-Z)}{Z}+\frac{2(1-Z)}{Z}\frac{\gamma_V}
{\sqrt{2\mu\varepsilon_B}}
\right).
\label{N27}
\ee

Now, for $|E_C|\gg\epsilon_B$, the effective-range expansion is
recovered from Eq.~(\ref{kcot}), with the scattering length $a$ and
the effective radius $r_e$ given by:
\be
a=\frac{2(1-Z)}{(2-Z)}\frac{1}{\sqrt{2\mu\varepsilon_B}},\quad
r_e=-\frac{Z}{(1-Z)}\frac{1}{\sqrt{2\mu\varepsilon_B}},
\label{weinberg}
\ee
which allows one to define the quantity $Z$ in a model-independent 
way, if the effective-range parameters $a$ and $r_e$ are available.

On the contrary, for $|E_C|\lesssim\epsilon_B$ the effective-range 
expansion does not converge anymore for energies larger than $|E_C|$. 
Correspondingly, formulae from Eq.~(\ref{weinberg}) do not hold
anymore. In the next section we discuss the circumstances, when
the described unusual behaviour may occur.

\section{Pole structure in the presence of the $t$-matrix zero}
\label{polestr}
 
Due to a very simple relation
(\ref{gf}) between $E_C$ and $\gamma_V$, it is instructive to study 
the amplitude singularities in terms of the variables $E_f$, $g_f$, and
$\gamma_V$. The expression for the scattering t-matrix in this case takes the form:
\be
t(E)=\frac{1}{4\pi^2\mu}
\frac{E-E_f+\frac12 g_f\gamma_V}{(E-E_f)(\gamma_V+ik)+\frac{i}{2}g_f\gamma_Vk}.
\label{tmat}
\ee
As seen from expression (\ref{tmat}), there 
generally could be up to three near-threshold poles, and
the equation defining the pole positions in the $k$-plane reads:
\be
(k^2-2\mu E_f)(\gamma_V+ik)+i\mu g_f\gamma_V k=0.
\label{eqk}
\ee
 
Denote the solutions of Eq.~(\ref{eqk}) as $k_1$, $k_2$, and $k_3$.
As follows from the explicit form of Eq.~(\ref{eqk}), one of these solutions is
always imaginary, while the other two are either 
both imaginary or placed at the lower half--plane 
symmetrically with respect to the imaginary axis, as required
by the general properties of the $S$-matrix. 
Thus, the set of parameters $\{E_f,g_f,\gamma_V\}$ used in Eq.~(\ref{tmat})
is fully defined by the poles of the scattering $t$-matrix $\{k_1,k_2,k_3\}$:
\begin{eqnarray}
E_f&=&-\frac{1}{2\mu}\frac{k_1k_2k_3}{k_1+k_2+k_3},\\
g_f&=&-\frac{(k_1+k_2)(k_1+k_3)(k_2+k_3)}{i\mu(k_1+k_2+k_3)^2},\\
\gamma_V&=&-i(k_1+k_2+k_3).
\end{eqnarray}

The position $E_C$ of the $t$-matrix 
zero can be also expressed in terms of $\{k_1,k_2,k_3\}$:
\be
E_C=E_f-\frac12 g_f\gamma_V=-\frac{1}{2\mu}(k_1k_2+k_1k_3+k_2k_3).
\label{ECfin}
\ee 
Notice that, for the given $E_f$ and $\gamma_V$, the range of values for $E_C$ is restricted by the condition 
$g_f\ge 0$ (see Eq.~(\ref{gf})), which follows from causality. 

It is illustrating to discuss the behaviour of the three poles when the
effective coupling $g_{f}$ is varied from zero to some large value. To
study the properties of the system we introduce the scale $\Delta\ll
\beta$ that defines the region of applicability of the 
equations\footnote{Note, however, that this applicability region might be further restricted by additional scales present in a particular problem under consideration.}.
If the coupling of the bare state to the mesonic channel is switched off, 
$g_f=0$, the solutions of Eq.~(\ref{eqk}) are
\be
k^{(0)}_{1,2}=\pm\sqrt{2\mu E_f},\quad k^{(0)}_3=i\gamma_V.
\label{bare}
\ee
The first pair of poles corresponds to the bare quark state (uncoupled
from the mesonic channel) with the
energy $E_f=E_0$, and the third one is due to the direct
interaction $V(\vep,\vep')$ in the mesonic channel, as 
already described above. 
Note, however, that the numerator of the $t$-matrix (see Eq.~(\ref{tmat}))
has a zero exactly at 
the momenta $k^{(0)}_{1,2}$ which eliminates the contribution of the corresponding 
poles.  
Thus, in the limit of an extremely small coupling, $g_f\ll|E_f|/|\gamma_V|$ 
(see Eq.~(\ref{ECfin})), there is only one pole $k_3$ in the scattering 
$t$-matrix that corresponds to a bound or virtual state
in the mesonic channel, depending on the sign of $\gamma_V$.
As the interaction $g_f$ is switched on and increases, $g_f\sim |E_f|/|\gamma_V|$,
the poles start to move and to mix, and the $t$-matrix may have a zero and three 
poles in different places of the $k$-plane. Note also that in this case 
there is no straightforward interpretation of $\gamma_V$
possible anymore due to the mixing of poles. Clearly, the zero of the 
$t$-matrix can  have an impact on the observables only if it is in the region 
of applicability defined as $\Delta$. 
From Eq.~(\ref{ECfin}) one can see then 
that $E_C \sim \Delta$ only if all three poles are located very close
to the threshold. In other words, if either the potential pole
or the quark poles are outside the region of applicability of our
formalism, also the $t$-matrix zero lies outside that region.  
In the latter case the effective-range
approximation is adequate for the problem under consideration, so that
one can set $E_C\gg E,\Delta$ in all equations, thus recovering the
standard Flatt{\'e} formulae.

Let us now assume that all three bare poles appear in the vicinity of
the threshold ($|E_f|\simeq\gamma_V^2/(2\mu)\lesssim\Delta$) 
but the coupling is strong ($g_f\gg |E_f|/|\gamma_V|$). 
Clearly, in this case $E_C \gg \Delta$ and  we again 
recover the standard lineshapes together with the standard
effective range expansion; only a single pole remains in
the near-threshold regime --- the expressions of Eq.~(\ref{weinberg})
hold again and we find, as expected, that this single pole corresponds
to a state of predominantly dynamical (=molecular) origin.

To summarise,
the most peculiar situation takes place if:
\begin{enumerate}
\item the bare quark and potential poles appear {\em accidentally} 
close to each other and to the threshold ($|E_f|\simeq 
\gamma_V^2/(2\mu)\lesssim\Delta$);
\item the quark state is relatively weakly coupled to the hadronic channel 
($g_f \sim |E_f|/|\gamma_V|$).
\end{enumerate}
Thus, the appearance of a zero in the scattering amplitude very close to
threshold, signaled also by an early breakdown of the effective-range
expansion, is a clear signal of the presence of both potential as well
as quark poles.

To detail these statements and to mimic the
peculiar situation described above we choose $|\gamma_V|\sim
\sqrt{2\mu |E_f|}$ for $g_f=0$. We choose for the reduced mass $\mu=1$~GeV as is
appropriate for the $X(3872)$ case, and define the near-threshold
region by $\Delta\sim 1$~MeV --- the energy range relevant for the 
$X(3872)$ particle (in case of the $X(3872)$ the range of validity
of the approach is not set by the range of forces, but by the
closet threshold, which is only 8 MeV away). 
We then study the system for different values
of $g_f$ for  six representative scenarios
(recall: $\gamma_V<0$ refers to an attractive potential,
which does not support a bound state, while $\gamma_V>0$
refers to a bound state present from the pure potential scattering).
For each case we fix the parameters $E_f$ and $\gamma_V$, and change 
$g_f$ from very small values to the ones for which the $t$-matrix zero 
leaves the near-threshold region. 

\begin{figure}[t]
\begin{center}
\epsfig{file=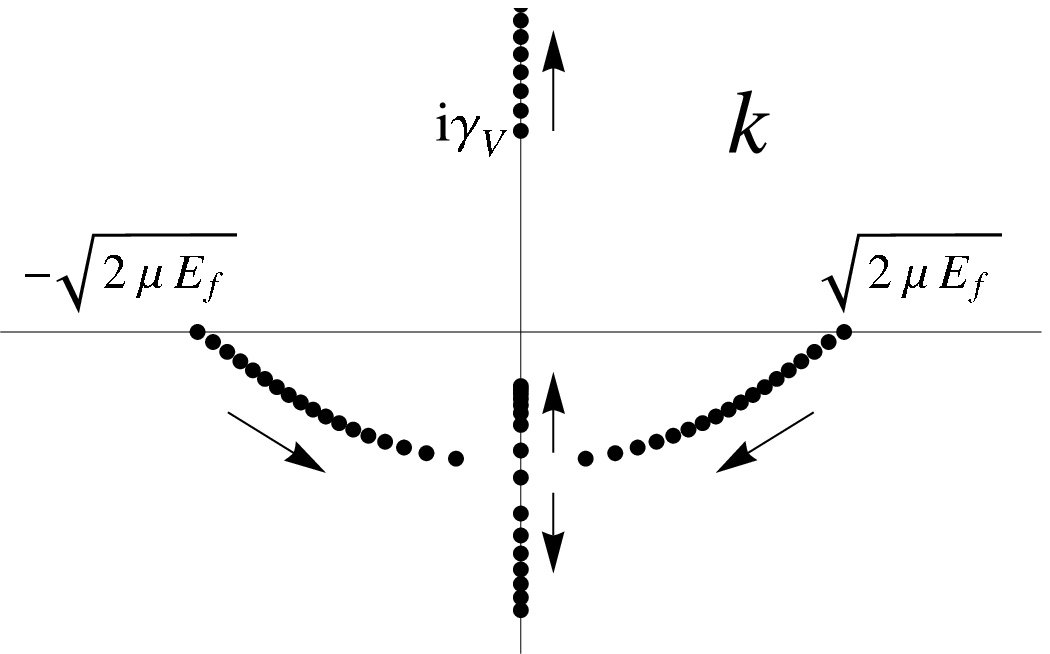,width=7.5cm}\hspace*{1cm}
\epsfig{file=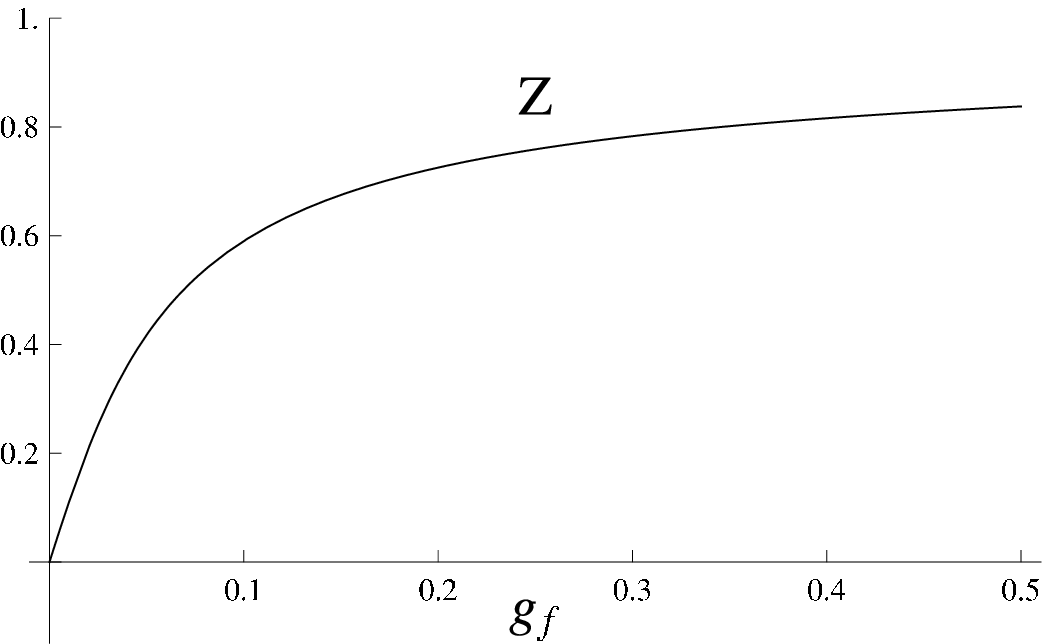,width=7.5cm}\\[5mm]
\epsfig{file=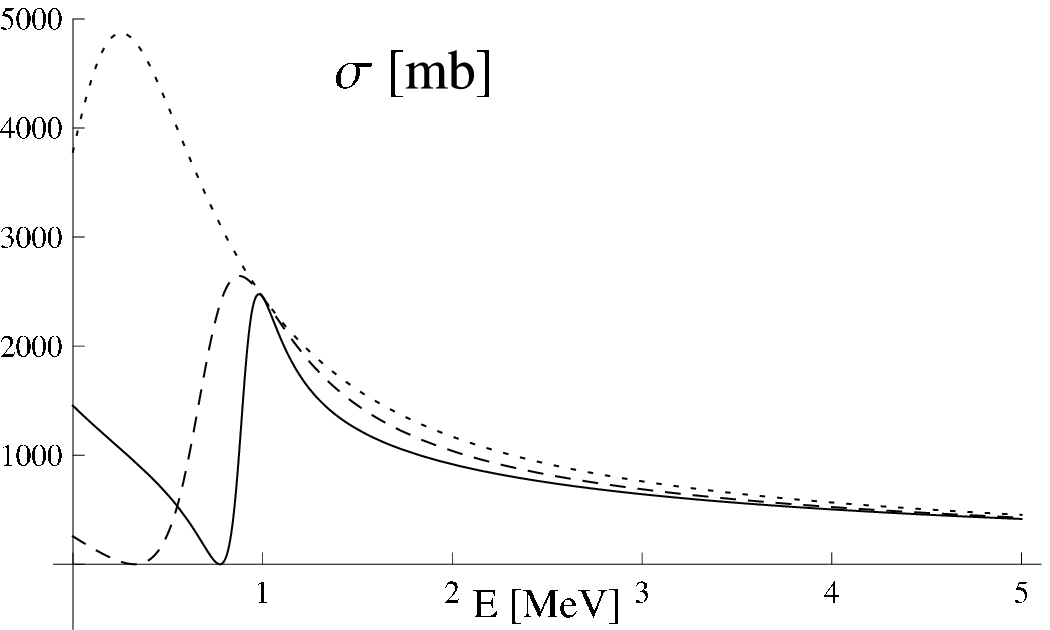,width=7.5cm}\hspace*{1cm}
\epsfig{file=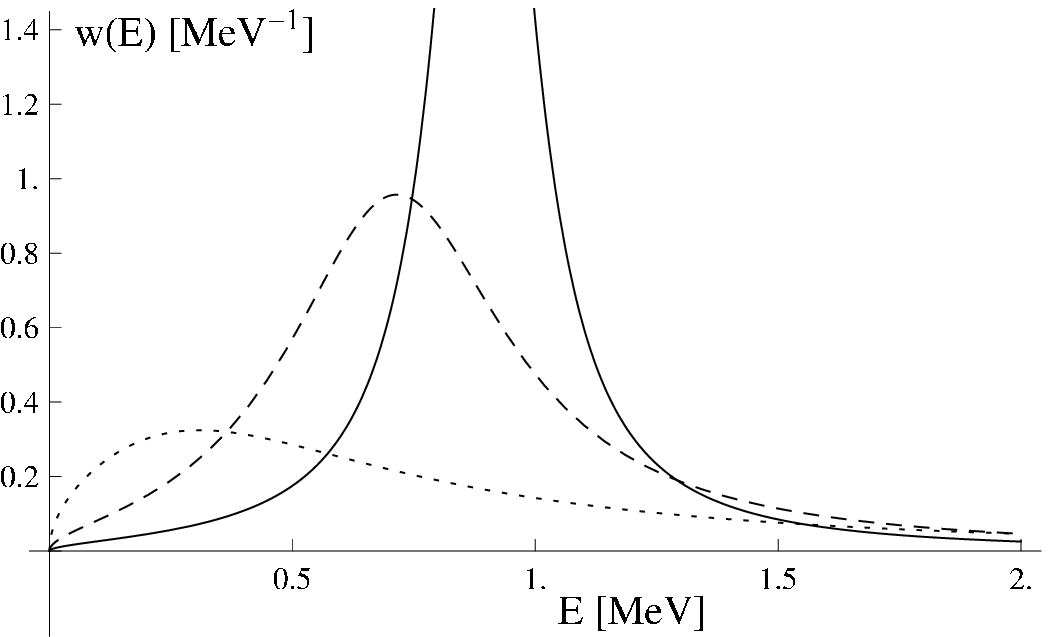,width=7.5cm}
\end{center}
\caption{Case (i). Upper panel: the pole structure in the $k$-plane (left
plot) and the $Z$-factor --- see Eq.~(\ref{Z}) (right plot) versus the coupling constant
$g_f$. Lower panel: the elastic scattering cross section (left plot) and the
spectral density (right plot) versus the energy for $g_f=0.01$ (solid line),
$g_f=0.03$ (dashed line), and $g_f=0.1$ (dotted line).}
\label{i-fig}
\vspace*{10mm}
\begin{center}
\epsfig{file=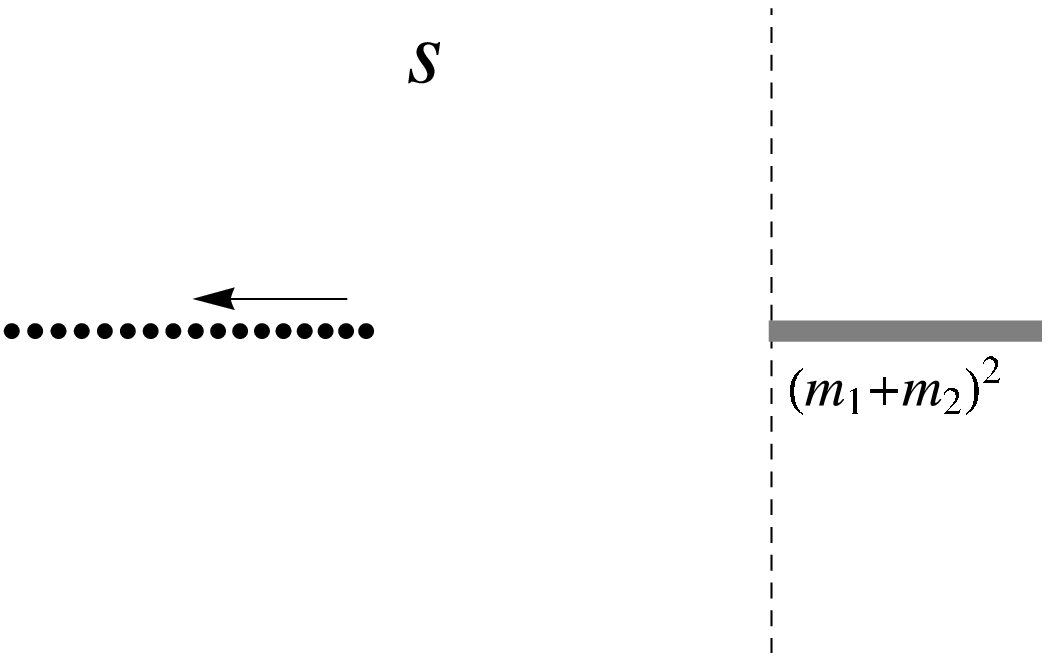,width=7.5cm}\hspace*{1cm}
\epsfig{file=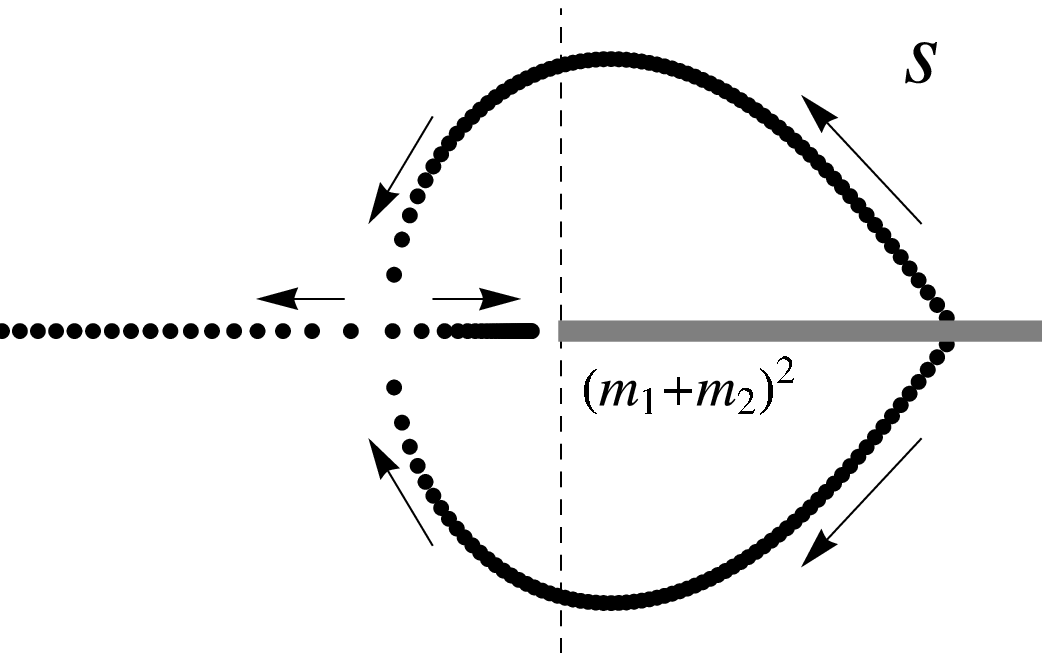,width=7.5cm}
\end{center}
\caption{The poles motion in the $s$-plane for Case (i) at the first Riemann sheet (left plot) and on the second
Riemann sheet (right plot). The unitarity cut is depicted in grey.}
\label{poles-fig}
\end{figure}

\begin{figure}[t]
\begin{center}
\epsfig{file=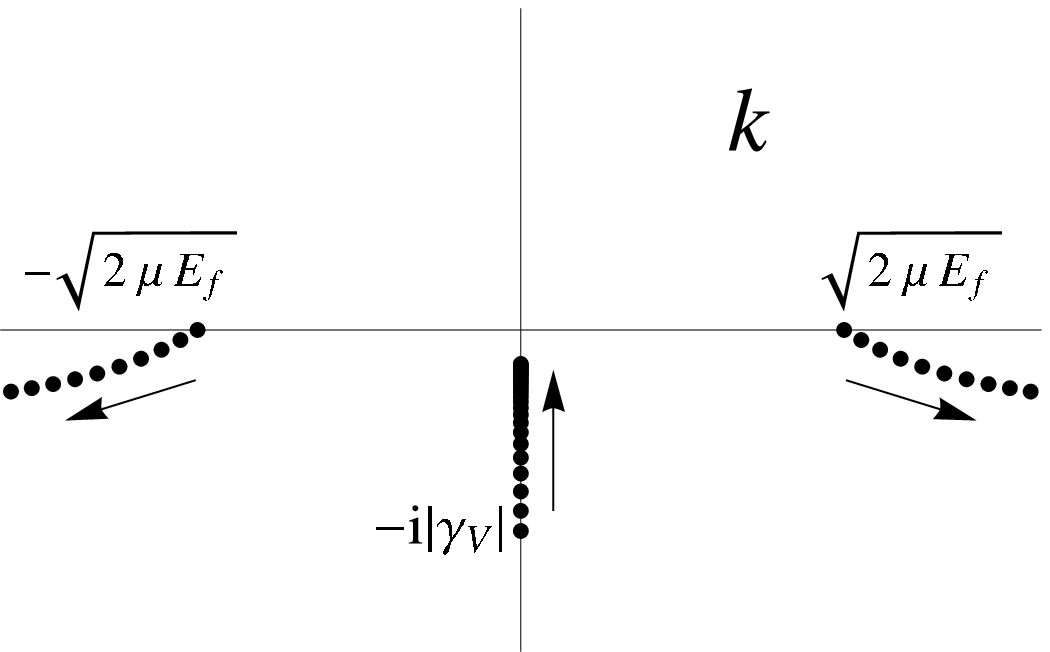,width=7.5cm}\\[5mm]
\epsfig{file=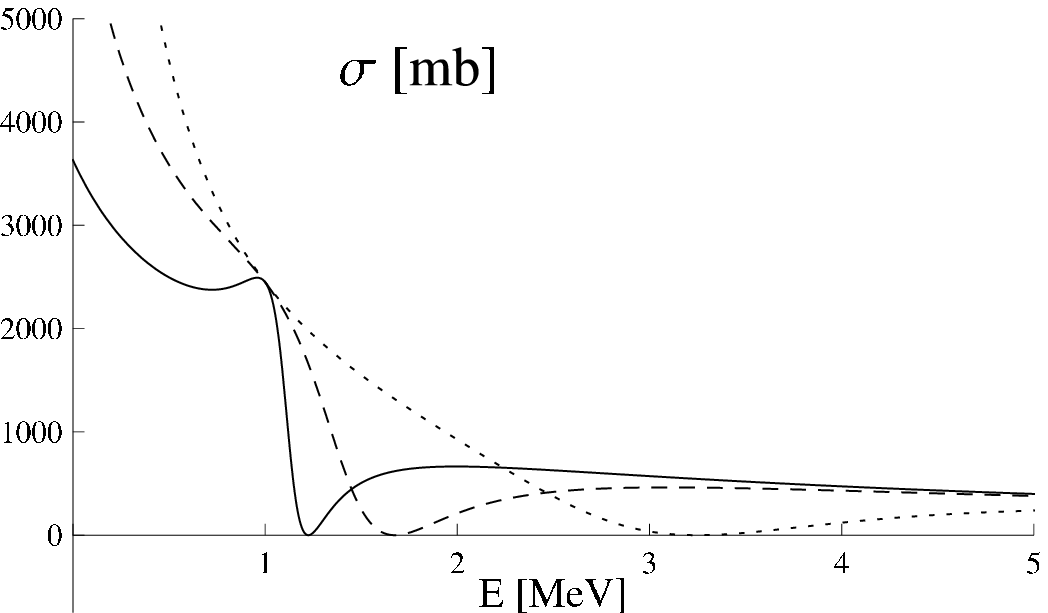,width=7.5cm}\hspace*{1cm}
\epsfig{file=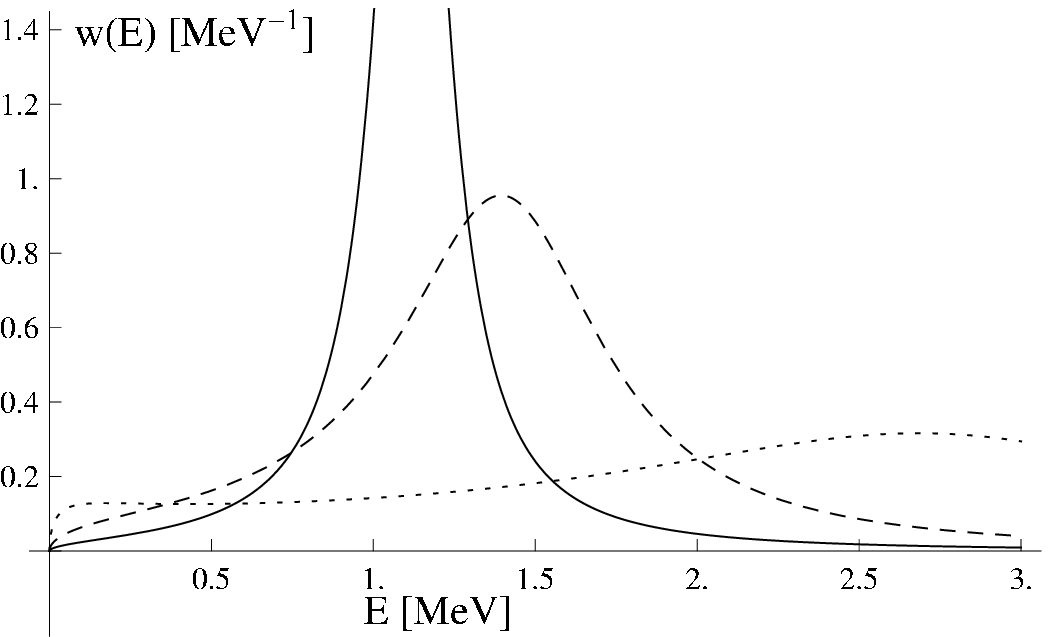,width=7.5cm}
\end{center}
\caption{Case (ii). Upper panel: the pole structure in the $k$-plane. Lower
panel: the elastic scattering cross section (left plot) and the
spectral density (right plot) versus the energy for $g_f=0.01$ (solid line),
$g_f=0.03$ (dashed line), and $g_f=0.1$ (dotted line).}
\label{ii-fig}
\vspace*{5mm}
\begin{center}
\epsfig{file=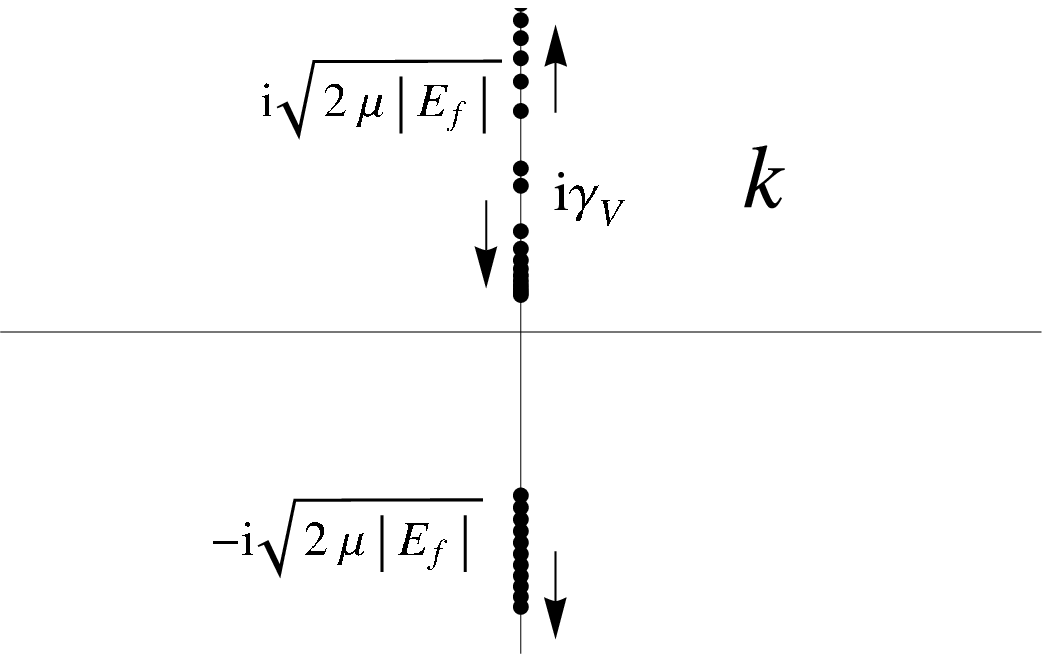,width=7.5cm}\hspace*{1cm}
\epsfig{file=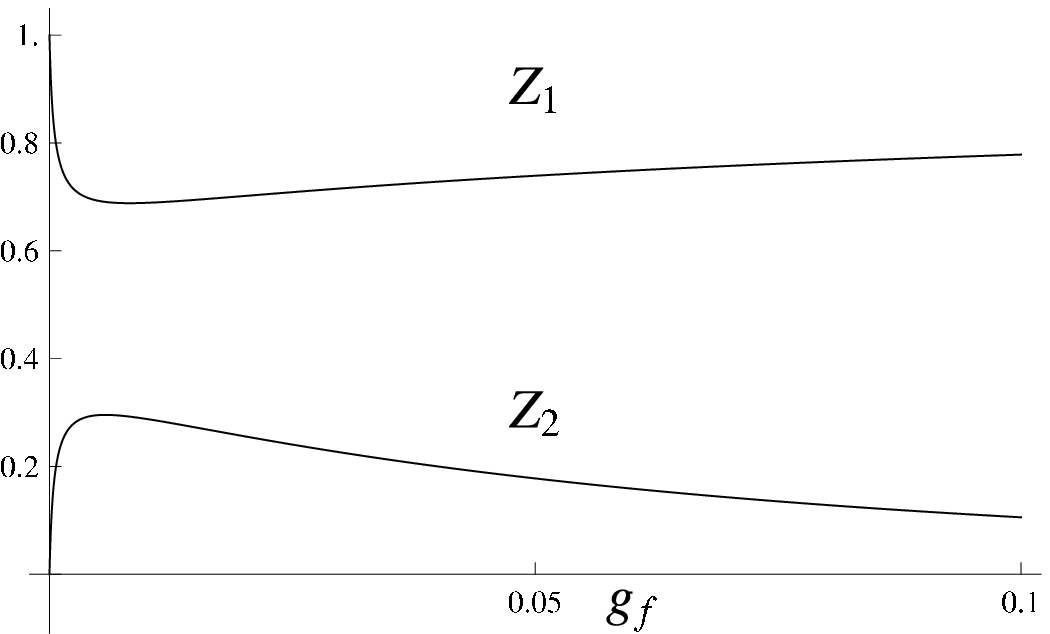,width=7.5cm}\\[5mm]
\epsfig{file=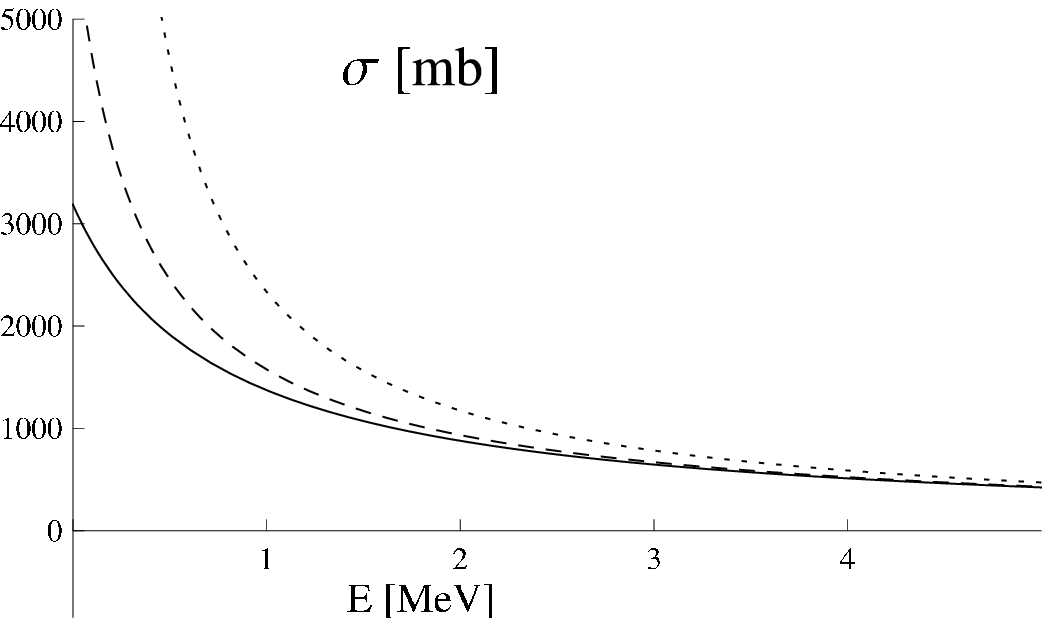,width=7.5cm}\hspace*{1cm}
\epsfig{file=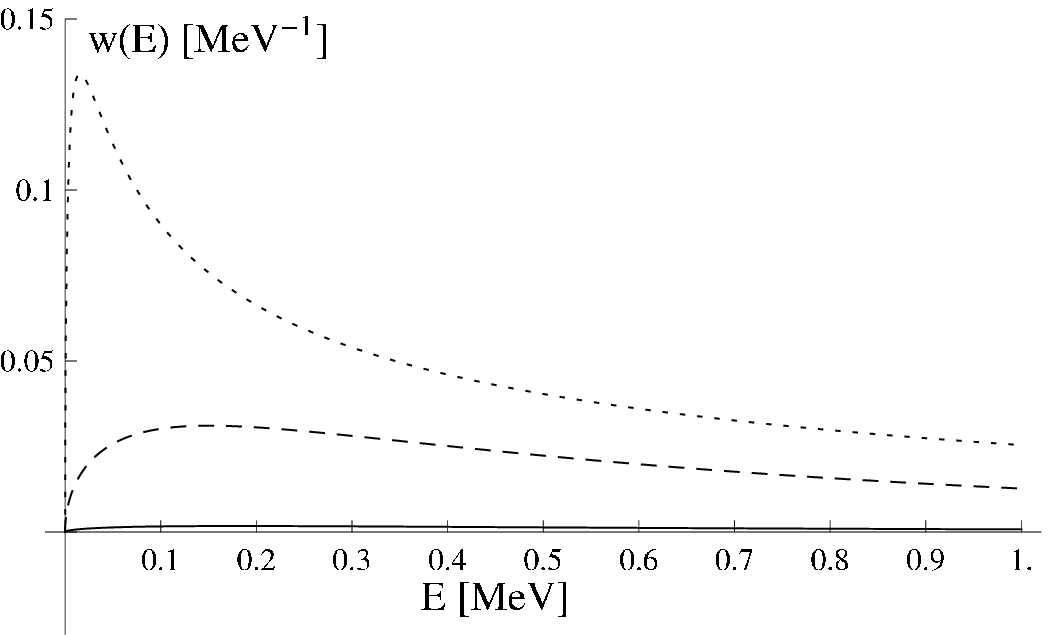,width=7.5cm}
\end{center}
\caption{Case (iii-a). Upper panel: the pole structure in the $k$-plane (left
plot) and the $Z$-factors --- see Eq.~(\ref{Z}) (right plot) versus the coupling constant
$g_f$. Lower panel: the elastic scattering cross section (left plot) and the
spectral density (right plot) versus the energy for $g_f=0.001$ (solid line),
$g_f=0.02$ (dashed line), and $g_f=0.3$ (dotted line).}
\label{iiia-fig}
\end{figure}

\begin{figure}[t]
\begin{center}
\epsfig{file=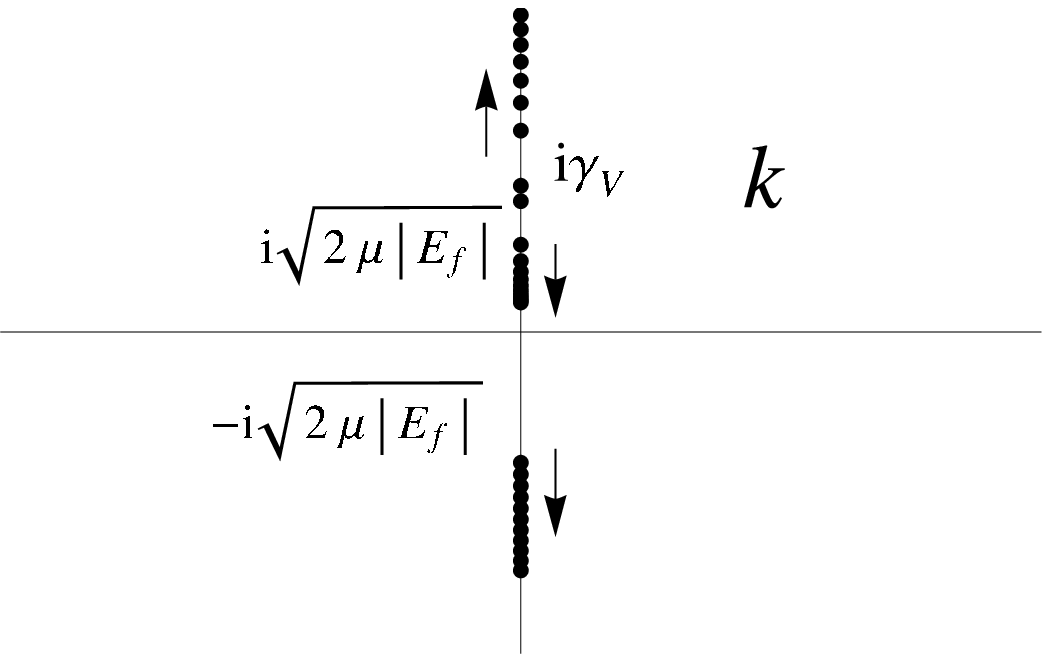,width=7.5cm}\hspace*{1cm}
\epsfig{file=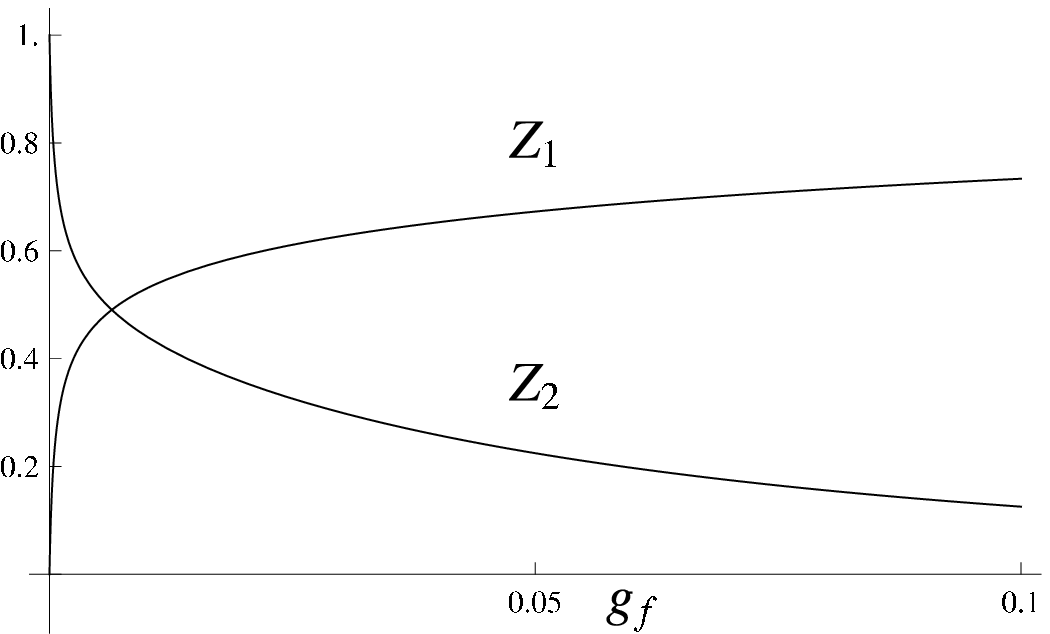,width=7.5cm}\\[5mm]
\epsfig{file=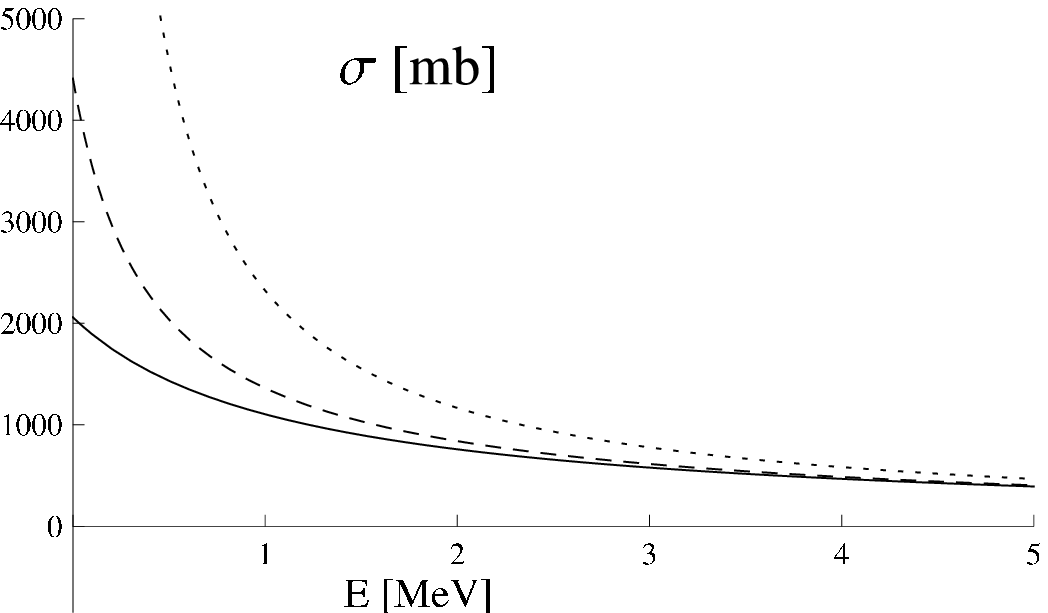,width=7.5cm}\hspace*{1cm}
\epsfig{file=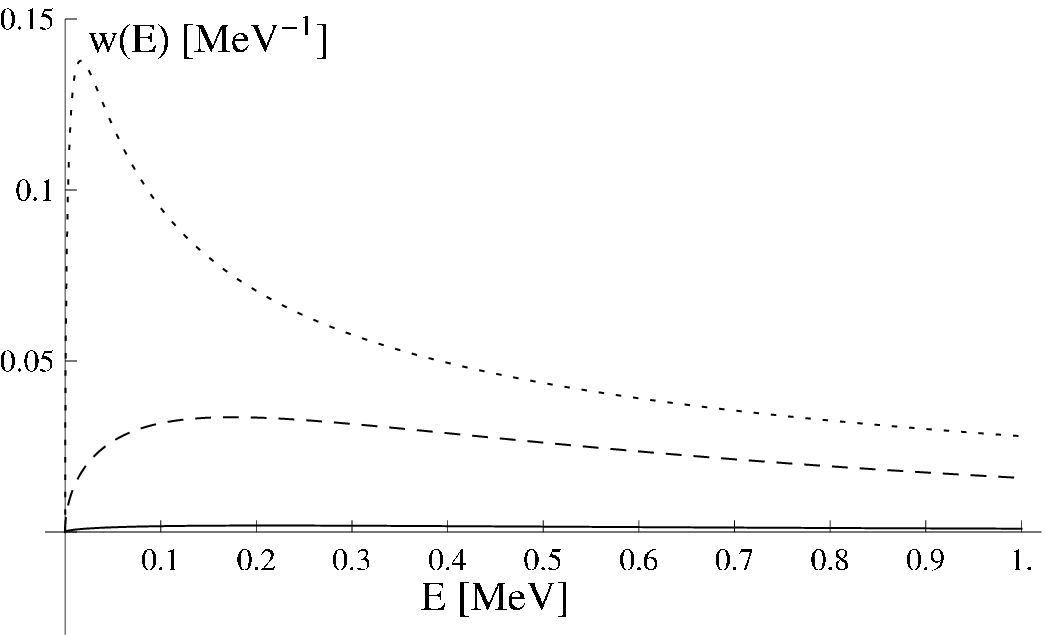,width=7.5cm}
\end{center}
\caption{Case (iii-b). Upper panel: the pole structure in the $k$-plane (left
plot) and the $Z$-factors (right plot) versus the coupling constant
$g_f$. Lower panel: the elastic scattering cross section (left plot) and the
spectral density (right plot) versus the energy for $g_f=0.001$ (solid line),
$g_f=0.02$ (dashed line), and $g_f=0.3$ (dotted line).}
\label{iiib-fig}
\vspace*{5mm}
\begin{center}
\epsfig{file=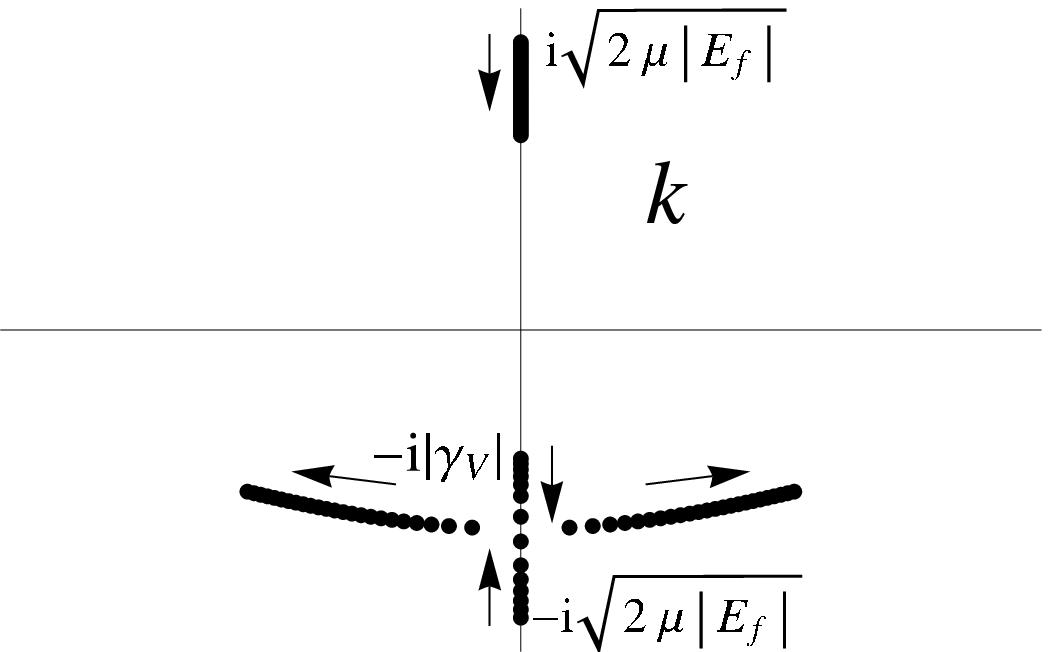,width=7.5cm}\hspace*{1cm}
\epsfig{file=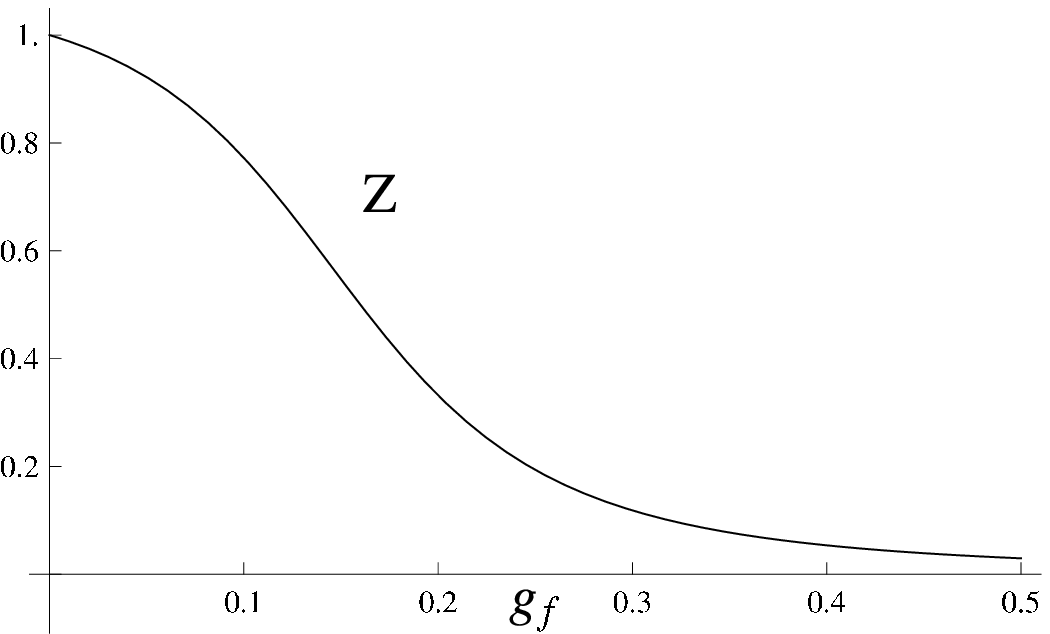,width=7.5cm}\\[5mm]
\epsfig{file=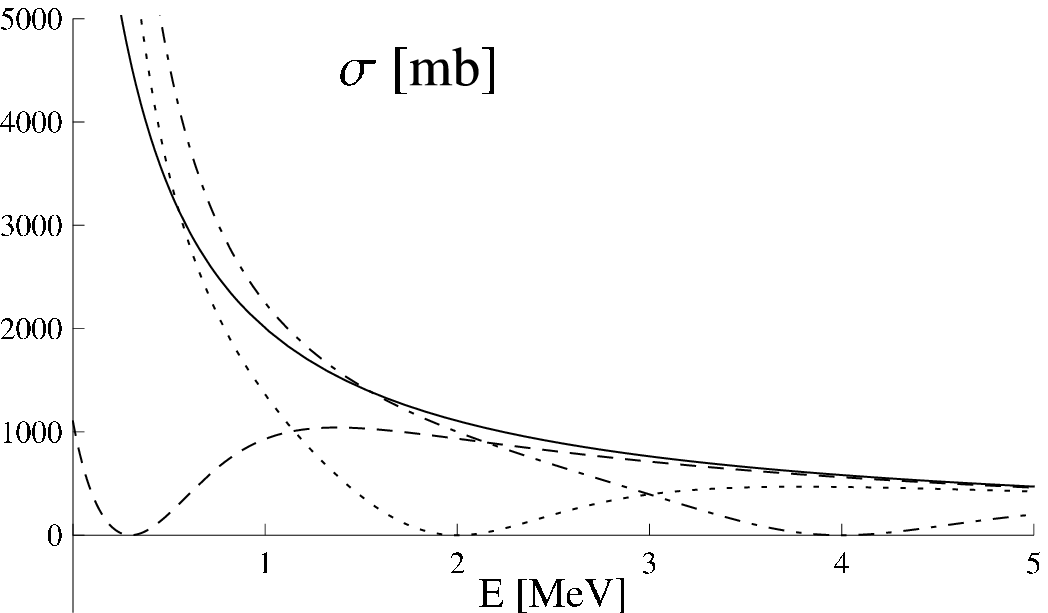,width=7.5cm}\hspace*{1cm}
\epsfig{file=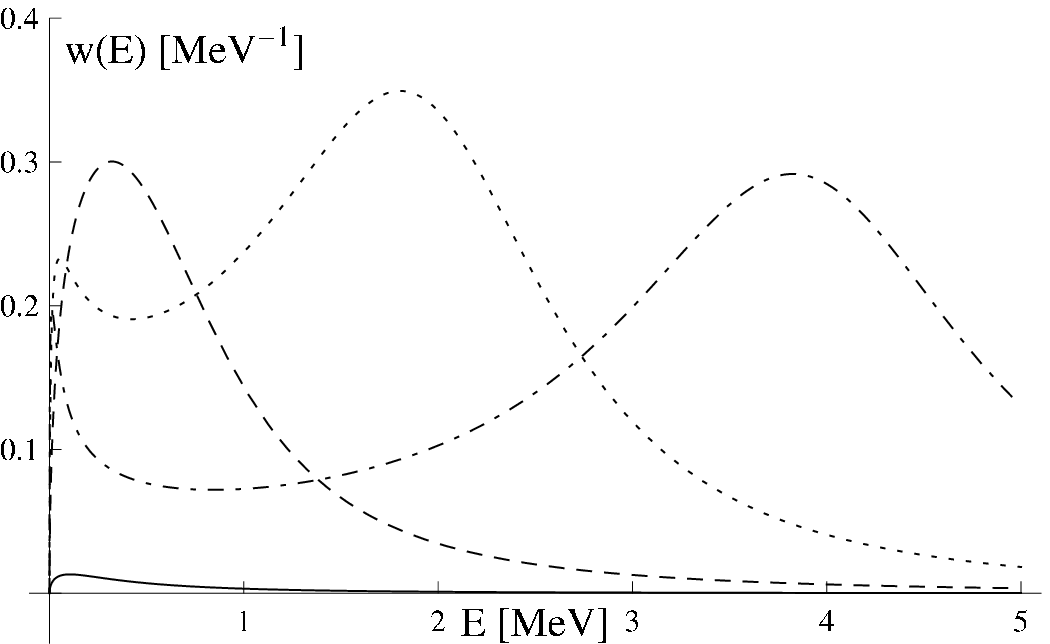,width=7.5cm}
\end{center}
\caption{Case (iv-a). Upper panel: the pole structure in the $k$-plane (left
plot) and the $Z$-factor --- see Eq.~(\ref{Z}) (right plot) versus the coupling constant
$g_f$. Lower panel: the elastic scattering cross section (left plot) and the
spectral density (right plot) versus the energy for $g_f=0.01$ (solid line),
$g_f=0.13$ (dashed line), $g_f=0.3$ (dotted line), and $g=0.5$ (dash-dotted line).}
\label{iva-fig}
\end{figure}

\begin{itemize}
\item Case (i): $E_f>0$, $\gamma_V>0$ ($E_f=1$~MeV,
$\gamma_V=45$~MeV). See Fig.~\ref{i-fig}, \ref{poles-fig}.
\item Case (ii): $E_f>0$, $\gamma_V<0$ ($E_f=1$~MeV,
$\gamma_V=-45$~MeV). See Fig.~\ref{ii-fig}.
\item Case (iii-a): $E_f<0$, $\gamma_V>0$, $\sqrt{2\mu 
|E_f|}>\gamma_V$
($E_f=-1$~MeV, $\gamma_V=40$~MeV). See
Fig.~\ref{iiia-fig}.
\item Case (iii-b): $E_f<0$, $\gamma_V>0$, $\sqrt{2\mu 
|E_f|}<\gamma_V$
($E_f=-1$~MeV, $\gamma_V=50$~MeV). See
Fig.~\ref{iiib-fig}.
\item Case (iv-a): $E_f<0$, $\gamma_V<0$, $\sqrt{2\mu 
|E_f|}>|\gamma_V|$
($E_f=-1$~MeV, $\gamma_V=-20$~MeV). See
Fig.~\ref{iva-fig}.
\item Case (iv-b): $E_f<0$, $\gamma_V<0$, $\sqrt{2\mu 
|E_f|}<|\gamma_V|$
($E_f=-1$~MeV, $\gamma_V=-55$~MeV). See
Fig.~\ref{ivb-fig}.
\end{itemize}

For cases (i) and (ii), the hadronic cross section vanishes at $E=E_C$
 accompanied by a very peculiar energy dependence. For
small couplings, the spectral density as well as the elastic cross
section peaks at around $E_f$ and the bound state, present in case
(i), is a purely mesonic one.  Then, with the increase of the
coupling, the aforementioned bound state leaves the near-threshold
region and acquires a large admixture of the quark component. In this
strong-coupling regime the dynamics is defined by the presence of a
single pole which corresponds to the virtual state.  Due to the
normalisation condition (\ref{wnorm}), the spectral density decreases
with the increase of the coupling.

It is instructive to discuss, for example, Case (i) in some more
detail.  In particular, in Fig.~\ref{poles-fig}, the movement of the
poles in the $s$-plane is depicted on both first (left plot) and
second (right plot) Riemann sheets. The arrows indicate how the poles
start to move when $g_f$ is increased. Clearly this plot is equivalent
to the pole movement in the $k$-plane depicted in the upper left panel
of Fig.~\ref{i-fig}. As usual, for a narrow resonance (small $g_f$)
only the pole in the lower half plane on the second sheet (see Fig.~\ref{poles-fig}) influences the physics. Here
this state is nearly a pure quark state. As $g_f$ gets larger the
width of the state grows (and the mesonic admixture increases). At
latest, when the real part of the pole position reaches the threshold
(indicated in the figure as the perpendicular, dashed line), both
poles are equally important.  Notice that, in this regime, the
interpretation of the imaginary part of the pole position as half of
the width of the state is lost. Similar pole movements in the
subthreshold regime were reported previously in
Refs.~\cite{withjose,withfengkun}.

In Cases (iii-a,b) one has $E_C<0$, so that the $t$-matrix zero
does not manifest itself in the hadronic cross section defined 
for positive energies only. In the meantime, with $|E_C|\sim \Delta$, 
there are two bound states in the near-threshold region.
With the increase of the coupling, only one bound state survives in the 
near-threshold region, with $Z \to 0$. 
The spectral density is small for all 
values of the coupling, as two bound states saturate the normalisation 
condition (\ref{wnorm}).

Finally, in cases (iv-a) and (iv-b) and for small couplings,
the bound state is almost purely of quark nature, and $E_C<0$. With the increase 
of the coupling, $E_C$ becomes positive and shows up in the 
hadronic cross section, while the bound state acquires a 
significant mesonic admixture. Further increase of the coupling 
forces the $t$-matrix zero to leave the near-threshold region, 
and the dynamics is defined by a single bound state, which is purely 
molecular, with $Z \to 0$.

So, for all cases considered, there is only one near-thresh\-old 
pole in the strong-coupling regime which corresponds to the 
scattering-length approximation for the mesonic $t$-matrix,
independently of the underlying dynamics. 

\section{Production reactions}
\label{prod}

\begin{figure}[t]
\begin{center}
\epsfig{file=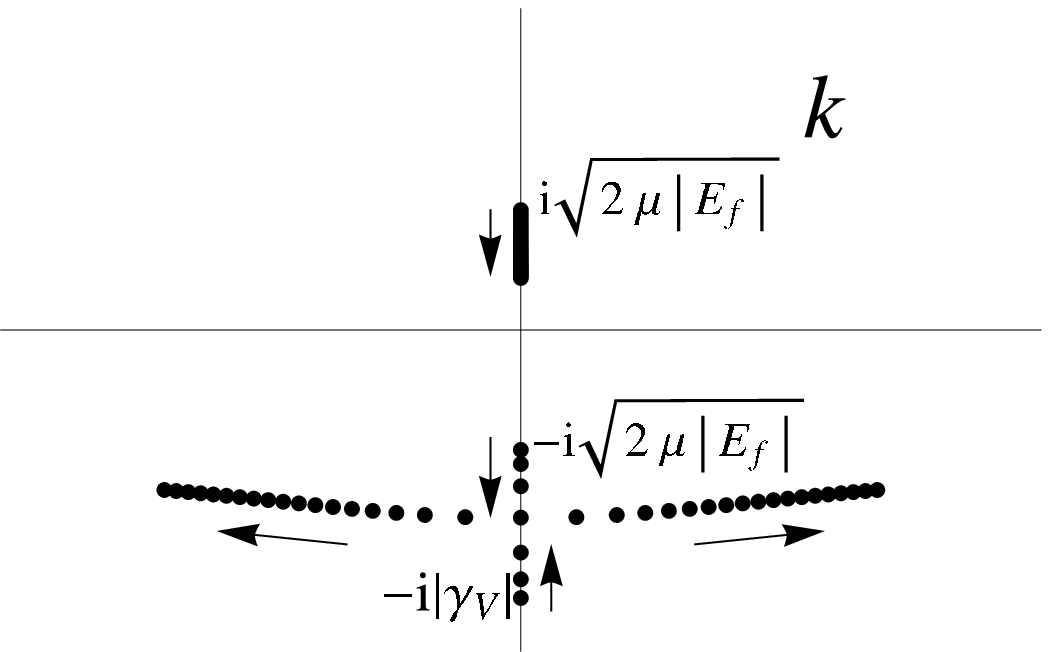,width=7.5cm}\hspace*{1cm}
\epsfig{file=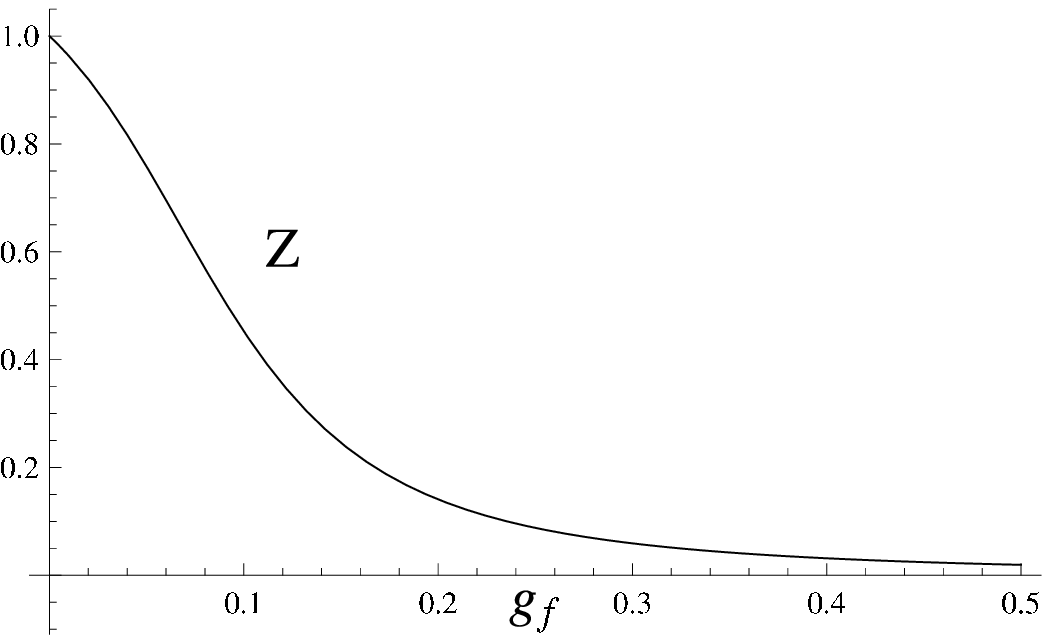,width=7.5cm}\\[5mm]
\epsfig{file=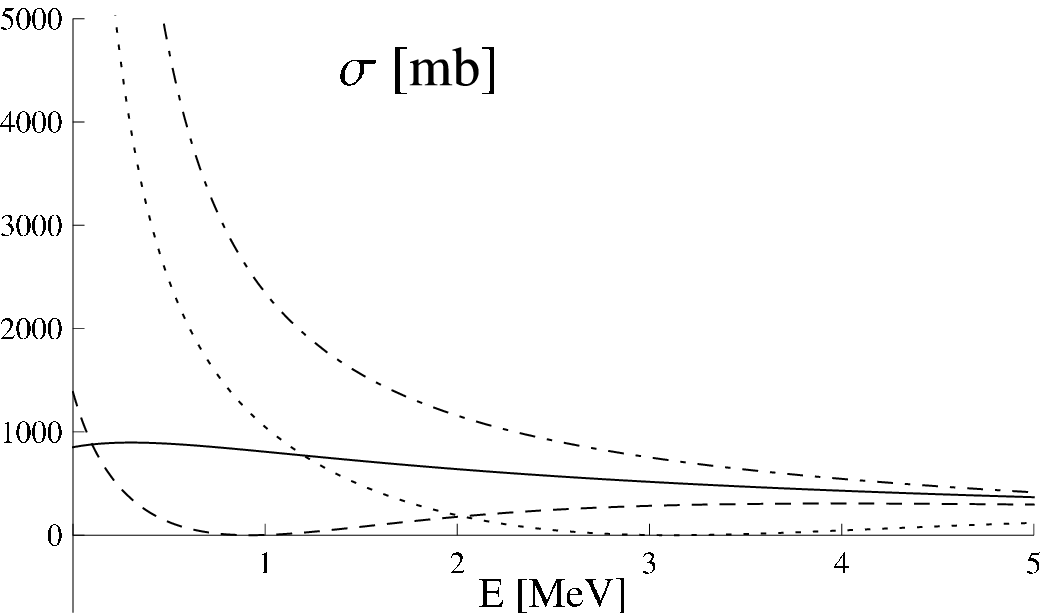,width=7.5cm}\hspace*{1cm}
\epsfig{file=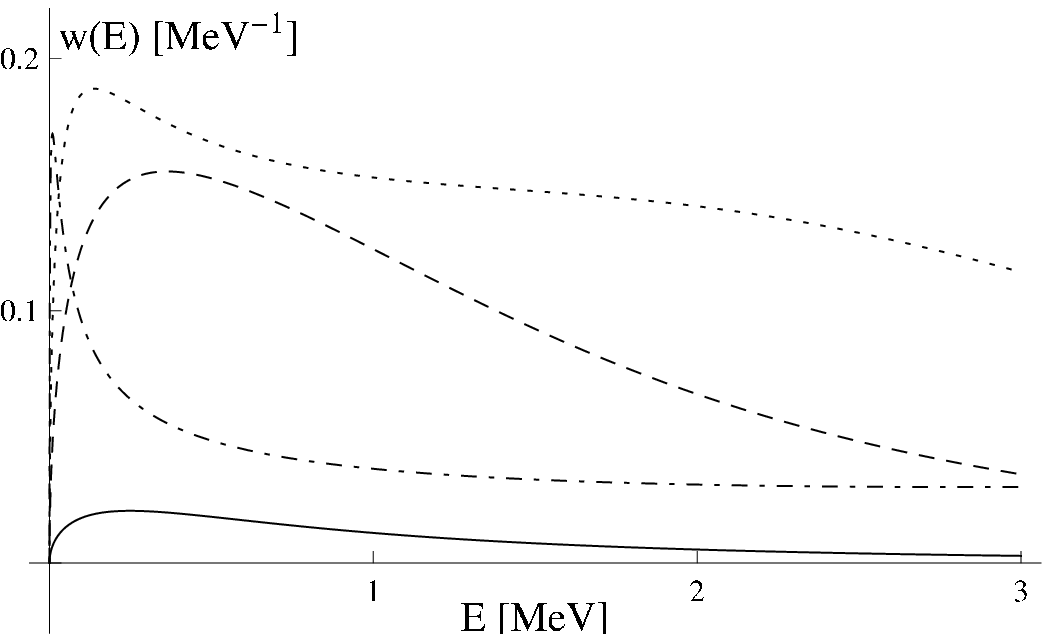,width=7.5cm}
\end{center}
\caption{Case (iv-b). Upper panel: the pole structure in the $k$-plane (left
plot) and the $Z$-factor --- see Eq.~(\ref{Z}) (right plot) versus the coupling constant
$g_f$. Lower panel: the elastic scattering cross section (left plot) and the
spectral density (right plot) versus the energy for $g_f=0.01$ (solid line),
$g_f=0.07$ (dashed line), $g_f=0.15$ (dotted line), and $g=0.5$ (dash-dotted line).}
\label{ivb-fig}
\vspace*{10mm}
\centerline{\epsfig{file=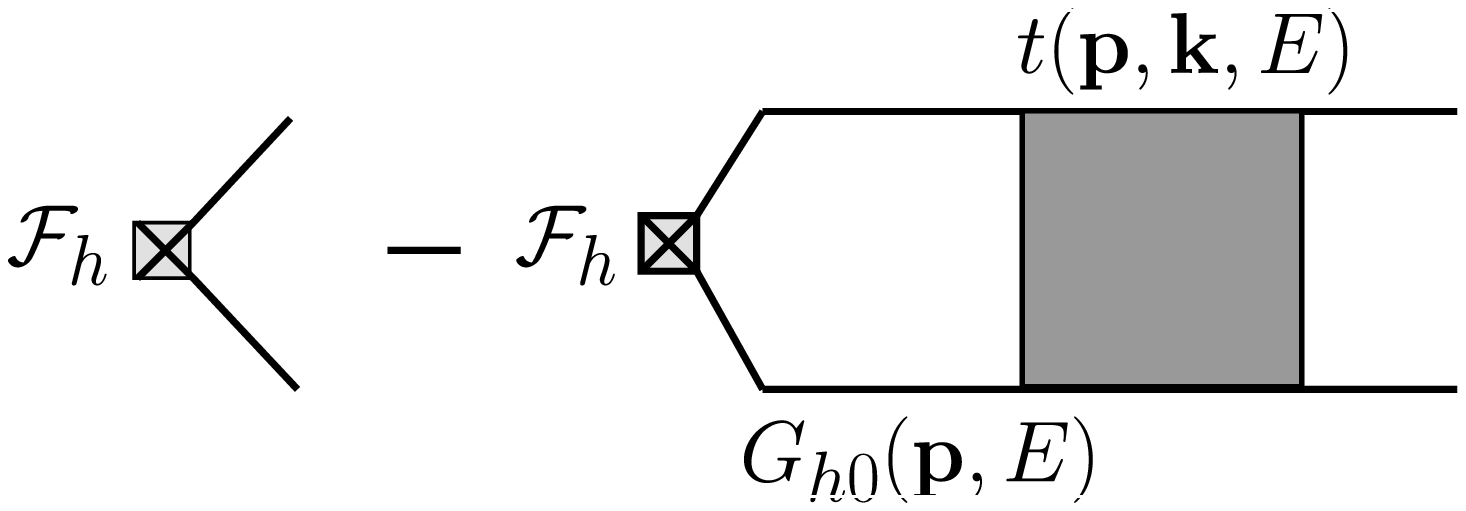,width=7.5cm}\hspace*{2.5cm}\raisebox{6mm}{\epsfig{file=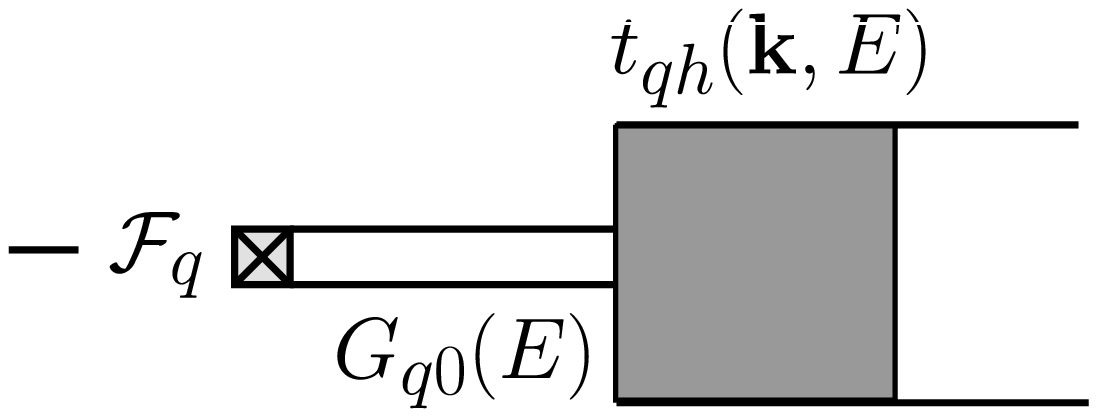,width=5cm}}}
\caption{Diagrams for the two meson production via the hadronic component (left plot) and via the
quark component (right plot).}\label{diag12fig}
\vspace*{10mm}
\begin{center}
\epsfig{file=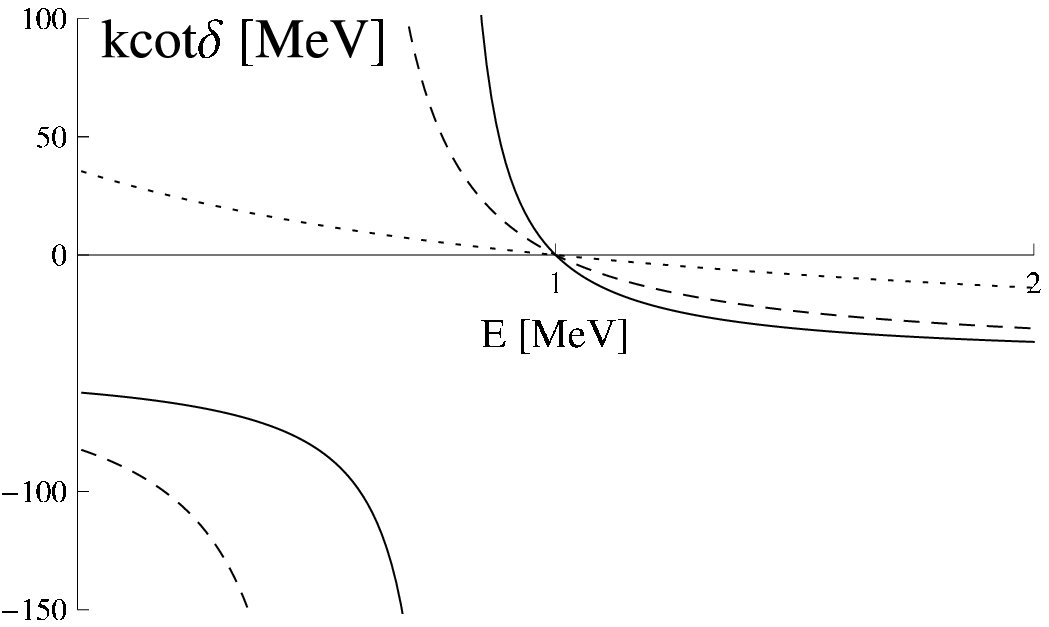,width=7.5cm}
\end{center}
\caption{The phase shift for Case (i)
for $g_f=0.01$ (solid line), $g_f=0.02$ (dashed line), and $g_f=0.1$ (dotted line).}
\label{ctg-fig}
\end{figure}

Unfortunately  there is no experimental possibility to study 
elastic scattering
of, say, charmed mesons, and our knowledge of the resonance
properties comes from production experiments.
In general there are two production mechanisms possible,
namely a production of the resonance of interest via its
hadronic component or via its quark component.
If the production
source can be considered as point-like, then the
production amplitude of the meson pair $(M_1M_2)$ through 
the resonance from the former mechanism
can be written as
\be
{\cal M}_h(E)={\cal F}_h\left(1-\int
d^3p\frac{t(\vep,\vek,E)}{p^2/(2\mu)-E-i0}\right)_{|k^2/(2\mu)=E},
\label{Mh}
\ee
where ${\cal F}_h$ is the initial state production amplitude from the
point-like source, unity
stands for the Born term, and the second term defines 
the final-state interaction --- see the left panel of Fig.~\ref{diag12fig}. For a 
point-like source and small energies, the amplitude (\ref{Mh}) is
\be
{\cal M}_h(E)={\cal F}_h(1-L(E)t(E)),
\ee
$L$ being the loop function describing
propagation of the intermediate mesonic state, 
\be
L(E)=4\pi^2\mu(l_0+ik),
\label{scale} 
\ee
with $l_0 \sim \beta$, where $\beta$ is the range of the force.
We may therefore write
\be
{\cal M}_h(E)={\cal F}_h\frac{(E-E_f)(E_f-E_C)-\frac12 l_0g_f(E-E_C)}
{(E-E_f)(E_f-E_C)+\frac{i}{2}kg_f(E-E_C)} \ .
\label{ph}
\ee
Thus, in ${\cal M}_h(E)$
the zero in the production amplitude is shifted with respect to 
the zero of the $t$-matrix. However, for the energies 
$E_f\simeq E_C\ll \beta$ this 
shift is small. The second term in the 
numerator of Eq.~(\ref{ph}) dominates, and the
production rate through the hadronic component alone is 
\begin{eqnarray}
\frac{dBr_h(M_1M_2)}{dE}&=&\mbox{const}\times 
k|t(E)|^2\Theta(E)\nonumber\\
&=&\mbox{const}\times 
\left(\frac{g_f}{8\pi^2\mu}\right)^2\frac{k}{|{\cal D}_F|^2} \ ,
\label{rh}
\end{eqnarray}
where the denominator ${\cal D}_F$ is given by Eq.~(\ref{Fden}).
Thus one might expect that all said above about the
physical content of a possible zero in $t$-matrix in the near-threshold
regime translates one-to-one also to production reactions.
However, this is not correct for there is, in addition to the
hadronic production, also the production via the quark component
possible. This piece can be expressed via the spectral density. Indeed,
as seen from the right panel Fig.~\ref{diag12fig}, the corresponding production amplitude is given by:
\be 
{\cal M}_q=-{\cal F}_qG_{q0}(E)t_{qh}(\vek,E),\quad\frac{k^2}{2\mu}=E, 
\ee
where ${\cal F}_q$ is the production amplitude for the bare quark
state by the point-like source, $G_{q0}(E)={1}/({E_0-E})$ is
the quark state free Green's function, and $t_{qh}(\vek,E)$ is the
$t$-matrix element responsible for the quark--meson transition. From the corresponding Lippmann--Schwinger equation,
\be
t_{qh}(\vek,E)=f(\vek)-\int d^3p \frac{f(\vep)t(\vep,\vek,E)}{p^2/(2\mu)-E-i0},
\ee
and with the help of Eqs.~(\ref{c}), (\ref{chi0}), the latter can be found in the form:
\be 
t_{qh}(\vek,E)=\int d^3p \chi_k(\vep)f(\vep)=(E-E_0)c_k(E).  
\ee
Thus, the near-threshold
production rate via the quark component of the wave function is simply
\be 
\frac{dBr_q(M_1M_2)}{dE}=\mbox{const}\times w(E),
\label{rq}
\ee
where $w(E)$ is defined in Eq.~(\ref{wE}). Then, using definition (\ref{wE}) and 
the explicit form of the coefficient $c_k(E)$, which follows from Eq.~(\ref{c}), 
\be
c_k(E)=\sqrt{\frac{g_f}{8\pi^2\mu}}\frac{E_f-E_C}{E-E_C}
\frac{1}{{\cal D}_F},
\ee
one finds:
\be
w(E)=\frac{1}{2\pi}\frac{kg_f}{|{\cal D}_F|^2}\frac{(E_f-E_C)^2}{(E-E_C)^2}.
\label{wf}
\ee
Obviously, this implies that $dBr_q/dE$ does not employ a zero. 

In reality one expects both mentioned mechanisms to contribute to the
production of the physical resonance, with a relative
importance depending on the production mechanism.
Therefore, we may write for the full production rate:
\be
\frac{dBr(M_1M_2)}{dE}=\mbox{const}\times k\;\frac{\left|E-E_C+r(E_f-E_C)\right|^2}{|{\cal
D}_F|^2(E-E_C)^2},
\label{full}
\ee where $r$ is a real number containing, among other contributions,
the ratio of the production rates via the quark state and the hadronic
state. A priori no estimate of $r$ is possible thus, even if there
were a system that shows a zero in the scattering amplitude near
threshold and allowing for all the conclusions of the previous
section, this zero might well be shielded in the production reaction.
On the other hand, if $E_c\gg \Delta$ it requires a delicate fine
tuning of the parameter $r$ to produce a zero in the amplitude of
Eq.~(\ref{full}). Thus, if there is a zero observed in a production
amplitude, it at least suggests the presence of both quark states as
well as a hadronic molecule.

Note that the zero in the lineshape of the $X(3872)$ predicted to occur in
a $B$-decay amplitude reported in Ref.~\cite{erics,voloshin3} is of a different
kind, for it originates as a coupled-channel effect of two nearby
continuum channels. In this work, on the other hand, the effects in the
presence of only a single continuum channel were discussed.

\section{Implications for lattice QCD}
  
From the previous discussions it should be clear that the
presence of a zero in the scattering amplitude contains 
important information about the system studied. However,
this information is no longer visible in the production 
amplitude, which is the quantity accessible experimentally.

Fortunately scattering observables, in particular phase
shifts, are accessible in lattice QCD from a study 
of volume dependencies
of the energy spectra \cite{luescher}. 
In simulations of full QCD both quark states as well
as molecular states are present and will influence
all correlators, as long as some overlap exists with
the interpolating fields used. Thus, the results of
those simulations need to be interpreted in the same
way as experimental results call for an interpretation,
if one wants to gain some insight into the dynamical 
mechanisms that lead to the structure formation. One
method pushed recently is to work with a large basis
of interpolating fields and to use the resulting eigenvectors
of the correlator matrices to interpret the findings~\cite{joe1,joe2}.
In Ref.~\cite{withfengkun} the scattering length is
shown to be an important quantity also to extract information
about the nature of states from lattice studies.
The discussion above provides one with an additional approach: in order to understand the interplay
of quark and meson degrees of freedom one may investigate
the near-threshold phase shifts. If a zero in the $t$-matrix,
corresponding to a zero in the phase shifts, occurs, it
immediately signals the presence of both quark poles
as well as potential scattering poles with a relatively weak
coupling as described in Sec.~\ref{polestr}. 

To illustrate this point we discuss briefly the
phase shift (its cotangent) for Case (i) introduced above,
\be
k\cot\delta=-\frac{\gamma_V(E-E_f)}{E-E_f+\frac12 g_f\gamma_V}.
\label{ctg}
\ee
This expression employs a pole at the position of the
$t$-matrix zero, as shown in Fig.~\ref{ctg-fig}. We expect
that such a structure could be extracted from studies
within lattice QCD.

\section{Conclusions}

In this paper we have presented a dynamical scheme in which a near-threshold zero in
the mesonic $t$-matrix appears as a consequence of the interplay
between quark states and a nonperturbatively interacting hadron--hadron continuum. The
appearance of such a zero invalidates the effective-range expansion
and corresponds to three near-threshold poles of the $t$-matrix. The
effect on hadronic observables in elastic scattering is proved to
be drastic.

However, a near-threshold $t$-matrix zero could exist  
only if several requirements are met. First, one needs the 
direct interaction in the mesonic channel to be strong enough to 
support a bound or virtual state. Second, a nearby bare quark state 
should exist, with a weak coupling to the mesonic channel. While, in 
principle, such a situation cannot be excluded, it is highly 
accidental. 

Without such special arrangements the effective-range 
formulae are valid, and the conclusions of Refs.~\cite{Weinberg1,Weinberg2,Weinberg3,evi1,evi2,Morgan} hold
true: a large 
and negative effective range corresponds to a compact quark state, while
a small effective range means that the state is composite. In the
former case there are two near-threshold poles in the $t$-matrix, 
while in the latter case there is only one pole.
  
In the two-pole case one can definitely state that the resonance is
generated by an $s$-channel diagram.  The coupling of this quark
state to the mesonic continuum is not large.  In the one-pole
situation no model-independent insight into the underlying dynamics is
possible and one can only indicate that the resonance is generated
dynamically but, in the language of meson exchange, the binding
potential could emerge from either $s$-channel of $t$-channel
exchanges --- or a mixture of both. It seems not possible to decide
between these scenarios model independently.  The three pole scenario,
on the other hand, calls for the presence of both $s$-channel
interactions as well as potential scattering --- traditionally
identified with $t$-channel exchanges. This situation can be
identified easily by the break-down of the effective-range expansion.
Although scattering experiments are not possible for most
unstable particles, an alternative access to scattering
observables could be provided by studies of lattice QCD. 

We also discussed the implications for production reactions.
Unfortunately it turns out that for the lineshapes of
production reactions no statement is possible
model independently based solely on information from the scattering 
amplitudes.

\begin{acknowledgement}
The work was supported in parts by funds provided from the Helmholtz
Association (grants VH-NG-222, VH-VI-231), by the DFG (grants SFB/TR 16
and 436 RUS 113/991/0-1), by the EU HadronPhysics2 project, by the RFFI (grants RFFI-09-02-91342-NNIOa and
RFFI-09-02-00629a), and by the Presidential programme for support of the leading scientific schools (grants
NSh-4961.2008.2 and NSh-4568.2008.2).
The work of V.B., Yu.K., A.K., and A.N. was supported by the State
Corporation of Russian Federation ``Rosatom''. A. N. would like to acknowledge the
support of the non-profit ``Dynasty'' foundation and ICFPM and of the grant PTDC/FIS/70843/2006-Fi\-si\-ca.
\end{acknowledgement}

\onecolumn

\end{document}